\documentclass[preprint,amssymb,aps,prb,superscriptaddress,floatfix]{revtex4}
\usepackage[pdftex]{graphicx}
\usepackage{color}
\begin{document}

\title{Role of molecular electronic structure in IETS: the
case of O$_2$ on Ag(110)}

\author{Serge Monturet}
\affiliation{Institut f\"ur Chemie, Universit\"at Potsdam, Karl-Liebknecht-Stra\ss e
24-25, D-14476 Potsdam-Golm, Germany}
\author{Maite Alducin}
\affiliation{Centro de F\'{\i}sica de Materiales Centro Mixto CSIC-UPV/EHU, Edificio
Korta, 20018 San Sebasti\'an, Spain}
\affiliation{Donostia
International Physics Center DIPC, P. Manuel de Lardizabal 4, 20018 San Sebasti\'an,
Spain}
\author{Nicol\'as Lorente}
\affiliation{Centre d'Investigaci\'o en Nanoci\`encia i Nanotecnologia (CSIC-ICN),
E-08193 Bellaterra, Spain}

\date{\today}
\begin{abstract}
Density functional theory (DFT) simulations corrected by the
intramolecular Coulomb repulsion $U$, are performed to evaluate the
vibrational inelastic electron tunneling spectroscopy (IETS) of O$_2$
molecules on Ag(110). Semilocal DFT calculations predict a spinless
adsorbed molecule, however the inclusion of the $U$ leads to the
polarization of the molecule by shifting a spin-polarized molecular
orbital towards the Fermi level. This has an important implication
in IETS, because a molecular resonance at the Fermi level can imply a
decrease in conductance while in the off-resonance case, an increase
in conductance is the expected IETS signal. We use the lowest-order
expansion on the electron-vibration coupling, in order to evaluate
the magnitude and spatial distribution of the inelastic signal. This
allows us to reproduce the experimental data [Hahn et al., Phys. Rev.
Lett. {\bf 85}, 1914 (2000)] in: (i) the negative conductance
variation observed in the vibrational spectra of O$_2$ along the
$[001]$ direction, (ii) the spatial distribution of the conductance
changes recorded over the O$_2$ molecule for the O--O stretch and the
antisymmetric O$_2$--Ag stretch vibrations, (iii) the absence of
signal for the center-of-mass and hindered rotations modes, and
(iv) the lack of IETS signal for the molecule chemisorbed along the
$[1\bar{1}0]$ direction. Moreover, our results give us insight of the
electronic and vibrational symmetries at play. The vibrational
frequencies need to go beyond the harmonic approximation in order to
be compared with the experimental ones, hence we present a
Morse-potential fitting of the potential energy surface in order to
evaluate accurate vibrational frequencies. The final IET spectra are
evaluated with the help of the self-consistent Born Approximation and
the effect of temperature and modulation-voltage broadening are
explored. This ensemble of results reveals that the IETS of O$_2$ is
more complicated that a simple decrease of conductance and cannot be
ascribed to the effect of a single orbital molecular resonance.
\end{abstract}

\maketitle

%Introduction
\section{Introduction}

Single-molecule vibrational spectroscopy is now possible thanks to the development of
the inelastic electron tunneling spectroscopy (IETS) operating with the scanning
tunneling microscope (STM)~\cite{Stipe}. The STM is thus conferred with the ability to
chemically analyze surfaces on the atomic scale~\cite{Ho,Komeda}. IETS has rapidly
become a mature technique and multimode analysis on molecular substrates have been
accomplished~\cite{Okabayashi} as well as analysis of extended vibrations or
phonons~\cite{Morgenstern}. The characterization and control of single molecule
reactivity through the excitation of a specific vibrational mode is a challenging area
for the STM-based IETS technique\cite{stprl97,ohkiprl08}. Within this context, the
interaction of O$_2$ with metals, which plays a central role in many technological
processes, has served as a model system to explore and control the fundamentals of
gas/surface reactivity.

The chemisorption and dissociation of O$_2$ on silver surfaces have
been widely investigated in last decades in an attempt to understand
the catalytic properties of silver, which are extensively exploited
in industrial-oxidation processes. Molecular beam experiments provide
detailed information on how reactivity depends on the collision
energy, the molecular rovibrational or electronic state, and the
surface temperature and coverage. Combining this technique with
either electron energy loss spectroscopy (EELS), infrared
spectroscopy, or thermal desorption spectroscopy (TDS), it is also
possible to extract information on the energetics ruling the
elementary gas/surface processes --activation energies, atomic and
molecular adsorption energies, reactive paths. Thanks to these kind
of studies we know that O$_2$ dissociation on flat Ag surfaces is
characterized by rather large activation
energies~\cite{Bird,albujcp08} and, as a consequence only molecular
adsorption is possible at crystal temperatures below
150~K~\cite{vaboprl94,buross96,rabuss96}. The chemisorption of O$_2$
on the Ag(110) surface has been particularly controversial because of
the initial disagreement regarding the nature and orientation of the
chemisorbed molecule. Density functional calculations performed by 
Gravil {\em et al.}~\cite{Bird} shed light to this controversy
showing the existence of two distinct chemisorption states with
essentially equal energies. This theoretical finding was subsequently
corroborated by EELS and TDS studies~\cite{bafrprl98}, though the
final confirmation came with the STM-IETS investigations by Hahn {\em
et al.}~\cite{Hahn,hahoprl01,JCP}. Compared with other
spectroscopies, the great advantage of IETS operated with STM is that
it provides simultaneous topographical and spectroscopical images with
atomic resolution. This permits an almost {\em direct} identification
of the molecular state.

Molecular O$_2$ chemisorbs on Ag(110) parallel to the surface over
the hollow site on two possible configurations, one with the
molecular axis along the $[001]$ surface direction (O$_2[001]$) and
the other one along the $[1\bar{1}0]$ direction (O$_2[1\bar{1}0]$).
Single molecule vibrations are detected with STM-IETS on the
O$_2[001]$ only. The recorded inelastic signal is a {\em decrease} in
conductance for one of the modes (the O--O stretch) and it can be a
{\em decrease} or an {\em increase} in conductance depending on the
exact position of the STM tip over the molecule for the other
detected mode (the antisymmetric O$_2$--Ag stretch)~\cite{Hahn}. The
reasons behind this rich and complex IETS structure have been a
matter of controversy that have recently been
solved~\cite{alsaprl10}.

In IETS, the changes of conductance are recorded as a function of the
tip-substrate voltage. It is generally admitted that in tunneling,
the excitation of a vibration mode leads to opening an inelastic
channel for conduction, hence increasing the
conductance~\cite{Persson,Lorente2000,Paulsson2,Tal}. However, early
model calculations predicted that when a molecular resonance overlaps
the Fermi level of the substrate, IETS could also give rise to
decreases in conductance~\cite{Persson}. Up to now, the
O$_2[001]$ is the only example, where such decreases in conductance
have been observed experimentally. Strikingly, state-of-the-art IETS
simulations, which use the electronic structure calculated with
DFT as the initial ingredient, have been unable to
reproduce the experimental conductance decreases. The simulations
give conductance increases because in the DFT electronic structure
the $\pi_g$ orbitals are not at resonance with the Fermi
level~\cite{Paulsson2,Lorente2004}. This discrepancy led to speculate
that the theoretical Fermi level was wrongly positioned.
However, questioning the position of the Fermi level and, thus,
the molecular charge state in this system is at variance with the
excellent characterization of the two experimental chemisorption
states given by this theory~\cite{Bird}. Moreover, the Fermi level
fitting, thought to force the $\pi_g$ resonance, cannot
reproduce the rich IETS structure of the experimental data showing
increases and decreases according to the vibrational mode and the tip
localization.

In a recent letter~\cite{alsaprl10}, it has been shown that
correcting the DFT electronic structure to adequately incorporate the
on-site Coulomb repulsion (the so called DFT+$U$ approach), the
experimental conductance changes recorded over the O--O stretch mode
and the antisymmetric O$_2$--Ag mode are correctly reproduced.
Furthermore, the DFT+$U$ calculations permit us to gain extra insight
on the actual electronic structure of this molecular system. Contrary
to the previous speculations, rather than shifting towards the Fermi
level, the $\pi_g$ resonance follows a Stoner-like splitting of its
spin components, leading to the partial filling of one of the spin
components of the $\pi_g$ orbitals and, thus, to the spin
polarization of the isolated O$_2$ molecule on Ag (110). In the
present contribution we present a comparative analysis of the DFT+$U$
electronic structure of the O$_2[001]$ and O$_2[1\bar{1}0]$ states,
discussing the properties that cause the respective survival and
quenching of the intrinsic O$_2$ paramagnetism. Our previous
simulations of the conductance changes for the O$_2[001]$ state are
extended to all the relevant molecule-surface modes and also to the
O$_2[1\bar{1}0]$ molecule. In particular, the reasons for the
existence or absence of IETS signal are explained in terms of the
electronic and vibrational mode symmetries. There are two symmetry
planes in the O$_2$/Ag(110) system. We show that the possibility to
excite a vibrational mode is determined by the overall symmetry
character of the coupling states and mode respect to those two
planes. However, the value of the IETS signal depends ultimately on
the energies of the orbitals and on the strength of the electron
vibration coupling. Also, we evaluate the IETS spectra with the help
of the self-consistent Born Approximation and explore the effects due
to temperature and modulation-voltage broadening.

The reminder of the paper is organized as follows.
Section~\ref{method-dft} describes the details of the DFT+$U$
calculations performed to obtain the electronic spectra of O$_2$
chemisorbed on Ag(110). In Sec.~\ref{method-modes}, we discuss the
necessity to go beyond the harmonic approximation in order to obtain
accurate vibrational frequencies of the molecule-surface modes. The
theoretical model used for the IETS simulations is explained in
Sec.~\ref{method-iets}. The calculated electronic structure is shown
in Sec.~\ref{result-dft}. The corresponding STM-IETS simulations are
discussed in Sec.~\ref{result-sym}. In Sec.~\ref{result-iets}, we
apply the SCBA model to calculate the IETS. Conclusions and final
remarks are presented in Sec.~\ref{con}.

\section{Methods}
\subsection{\label{method-dft} Details of the DFT+$U$ calculations}
%DFT details
Density functional theory calculations are performed with the
{\footnotesize VASP} code~\cite{vasp} using plane waves with an
energy cutoff of 515~eV and the projector augmented wave
method~\cite{paw}. We use a periodic supercell consisting of a
six-layer slab separated by 10.45~\AA~of vacuum and a (3$\times$4)
surface unit cell large enough to significantly reduce interactions
among oxygen molecules in neighboring cells. The surface Brillouin
zone is sampled with a 4$\times$4$\times$1 Monkhorst-Pack grid of
special $k$ points. The exchange and correlation energy is calculated
within generalized gradient approximation (GGA) using the PW91
functional~\cite{pw91}. Corrections to the on-site Coulomb repulsion
are applied to the oxygen 2$p$ electrons following the rotationally
invariant DFT+$U$ scheme proposed by Dudarev {\em et
al.}~\cite{duboprb98} as implemented in {\footnotesize VASP}. The
screened on-site Coulomb interaction $U$ is calculated from first
principles as the energy cost for adding extra charge to O$_2$ when
adsorbed on Ag(110). The value obtained with the method described in
Ref.~[\onlinecite{cogiprb05}] is close to 4~eV. However, we also
verify that the results shown in this work, are reproduced for $U$
varying within the range 2--4~eV. Such a verification is necessary
because the calculated $U$ value actually depends on the projectors and
the used constrained method to fix the final electronic structure.
The adsorption positions are optimized by fully relaxing the O atoms
and the two uppermost silver layers until atomic forces are less than
0.02~eV/\AA.

\subsection{\label{method-modes}Vibrational frequency evaluation: beyond
the harmonic approximation}

The diagonalization of the dynamical matrix gives the vibrational modes within the
harmonic approximation. The method implemented in {\footnotesize VASP} uses finite
differences to compute the second derivative of the hamiltonian with respect to a
constant displacement $\Delta R$ of every coordinate of the active atoms, here the
oxygen ones. In Table~\ref{table_modes} we give the first 4 molecular modes on the
surface for atomic displacements, $\Delta R=$ 0.025, 0.03 and 0.04~\AA . The
displacement of 0.04~\AA~is the one that has the better overall agreement with the
experimental measurements. The dispersion is still considerable, despite DFT is known
to yield frequencies within $\sim 5$\% the experimental values~\cite{Martin}. The use
of a rigid displacement to estimate the dynamical matrix is too simple to capture the
complexity of molecular modes on surfaces. Indeed, a small displacement will be probing
a region too flat of the potential energy surface (PES), and a large value will be
probably in the anharmonic part of the PES. The situation becomes more difficult when
we consider modes that span several tens of meV as is the present case. A $\Delta R$
that probes the harmonic potential corresponding to the center-of-mass coordinates (at
$\sim 30$ meV) will be too large to probe the O--O stretch mode (at $\sim 80$ meV).

\begin{figure}
\suppressfloats
%\begin{center}
%\includegraphics*[width=0.75\columnwidth]{figure0.prb.eps}
\includegraphics*[width=0.75\columnwidth]{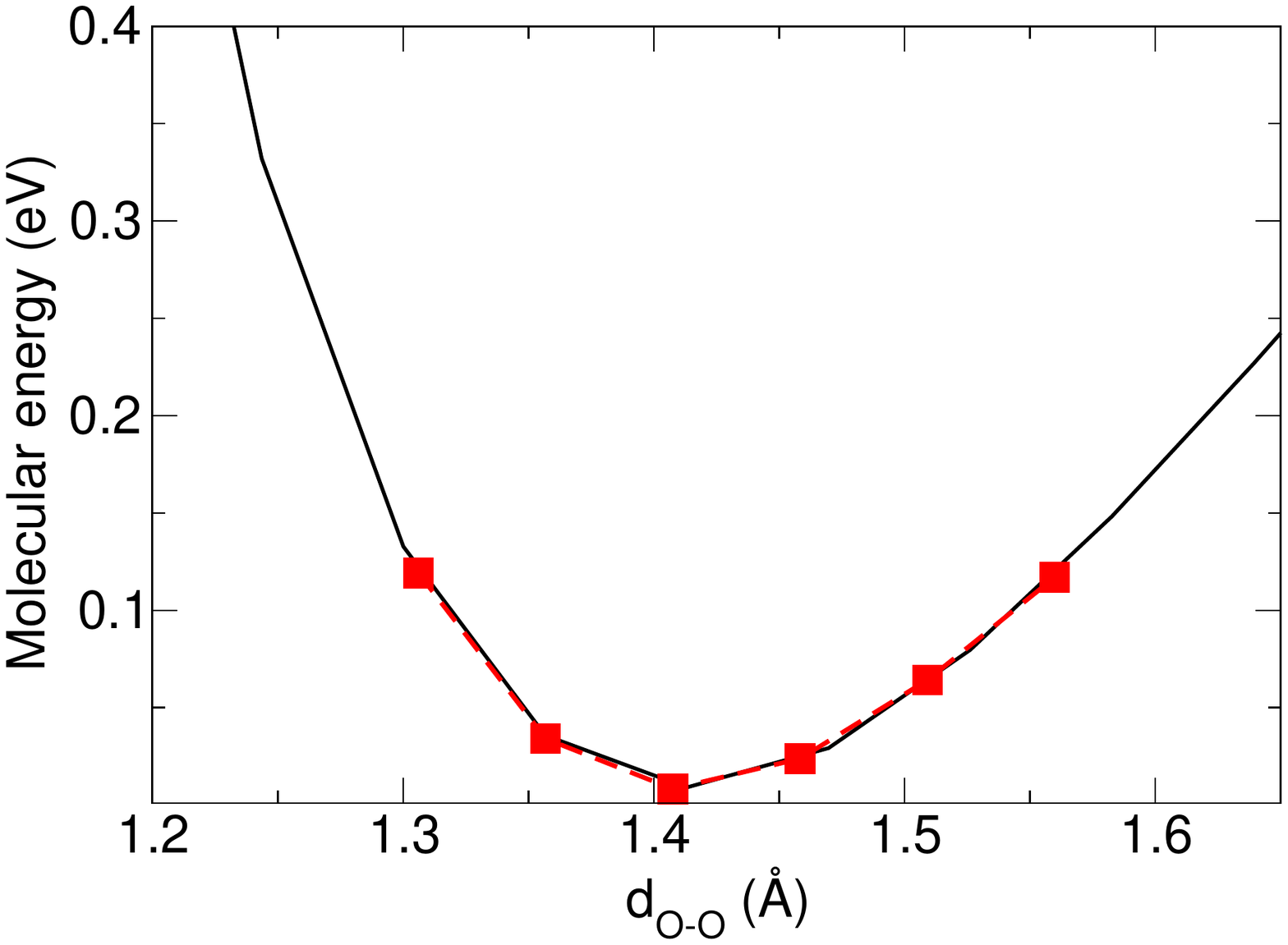}
%\end{center}
\caption{ Potential energy curve along the O--O stretch mode of O$_2$ on Ag(110)
(squares) and a Morse potential fit (black line). The obtained frequency is 83.6 meV in
excellent agreement with the experimental one (82.0 meV)~\cite{Hahn}.} \label{morse}
\end{figure}

The vibrational frequencies are accurately calculated
%Very good agreement is however obtained
if several points of the PES are obtained and fitted to a Morse
potential. This is the strategy followed here. We use the
eigenvectors of the simple finite difference method. These
eigenvectors change little from displacement to displacement (less
than the frequencies do) and are a good starting point to plot the
PES along a given mode. Figure~\ref{morse} shows the PES along the
O--O stretch mode and the corresponding Morse-potential fit.  Solving
the one dimensional the Schr\"odinger equation for this PES, we find
a frequency of 83.626 meV. This value is now in good agreement with
the experimental frequency of 82.0 meV. Therefore, we safely conclude
that the uncertainty in the frequency evaluation is due to the
anharmonicities of the PES and not to the DFT calculation.

\subsection{\label{method-iets} IETS simulations}

The IETS simulations of O$_2$ on Ag(110) are performed using two different methods. For
the quantitative evaluation of IETS and its spatial distribution we use the many-body
extension of the Tersoff-Hamman theory for the STM~\cite{Lorente2000,Lorente2004}. The
current implementation is based on DFT results regarding the electronic structure and
the electron-vibration couplings. Briefly, the conductance change due to the excitation
of a localized vibration to the lowest order in the electron-vibration coupling, $v$ is
given by the opening of the inelastic channel, leading to an increase of conductance,
$\Delta \sigma_{ine} ({\bf r})$, and the change in the elastic channel to the same
order in $v$, $\Delta \sigma_{ela} ({\bf r})$, which means a decrease in conductance.
In Ref.~[\onlinecite{Monturet08}], an all-order theory shows that the decrease of
conductance is related to the appearance of vibrational side-bands in the elastic
electron transmission. The decrease is maximum when a molecular level is resonant with
the Fermi energy and such that side bands appear at $\pm \hbar \Omega$.

Due to the strong molecular character at the Fermi energy,
approximations replacing the energy dependence by the behavior at the
Fermi energy are not correct any longer. Hence, we use the full
electronic structure and energy dependence for the finite bias of the
measurements. In this case, we extend the customary IETS
treatment~\cite{Lorente2000,Lorente2004} to new equations that keep
all the energetic dependence. Hence, the inelastic contribution to
the conductance change is given by:
\begin{eqnarray}
\frac{\Delta \sigma_{ine}}{\sigma} &=& \frac{1}{\rho(\mathbf{r}_0, E_F+eV)}\;
\nonumber \\
&\times& \sum\limits_{n, \mathbf{k}} \left| \sum_{m} \; \frac{\left \langle \psi_{m,
\mathbf{k}}|v|\psi_{n, \mathbf{k}}\right \rangle \psi_{m, \mathbf{k}}
(\mathbf{r}_0)}{\epsilon_{n, \mathbf{k}}-\epsilon_{m, \mathbf{k}}+i 0^+} \right|^{2}\;
(1-f(\epsilon_{n, \mathbf{k}}))\; \delta(E_F + eV -\hbar \Omega - \epsilon_{n,
\mathbf{k}}) \, , \label{ine}
\end{eqnarray}
where $\rho(\mathbf{r}_0, E_F+eV)$ is the local density of states evaluated at the STM
tip's center $\mathbf{r}_0$ and at the Fermi energy plus the energy of the electron at
the corresponding bias, $V$. The Fermi distribution function is given by $f(\epsilon)$
and $\Omega$ is the frequency of the considered mode. Since we are using
periodic-boundary calculations, the electronic states are Bloch states with band
indexes $n$ and $m$ and k-vector $\mathbf{k}$. Note that the local electron-vibration
potential $v$ couples states with the same k-vector. The elastic term is
\begin{eqnarray}
\frac{\Delta \sigma_{ela}}{\sigma} &=&
\frac{-2 \pi^2}{\rho(\mathbf{r}_0, E_F+eV)} \; \nonumber \\
&\times &\sum\limits_{n, \mathbf{k}} \left| \sum_{m} \; (1- f(\epsilon_{m,
\mathbf{k}})) \left \langle \psi_{m, \mathbf{k}}|v|\psi_{n, \mathbf{k}}\right \rangle
\psi_{m, \mathbf{k}} (\mathbf{r}_0) \; \delta (\epsilon_{m, \mathbf{k}}- \hbar \Omega
-\epsilon_{n, \mathbf{k}})
\right|^{2} \; \label{ela}\\
&\times&(1-f(\epsilon_{n, \mathbf{k}})) \; \delta(E_F + eV -\hbar \Omega - \epsilon_{n,
\mathbf{k}}) \, . \nonumber
\end{eqnarray}
The factor $(1-f(\epsilon_{n, \mathbf{k}}))\delta(E_F + eV -\hbar
\Omega - \epsilon_{n, \mathbf{k}})$ is responsible for the
temperature-dependent onset of the vibrational signal. Indeed, this
factor is strictly zero at zero temperature, if $eV < hbar \Omega$.
In the evaluation of these equations the k-point sampling is critical
because a large number of electronic states is needed to ensure
numerical accuracy. The $\delta$-functions are approximated by
Gaussian functions of broadening smaller than the used bias.
Typically, we have performed the simulations at $V=150$~mV with
a gaussian broadening of 50~meV, except for the O--O stretch mode that is calculated
at $V=200$~meV. The experimental change in conductance is simulated
as the sum of the inelastic and elastic contributions.

The second method is the use of the self-consistent Born approximation
(SCBA)~\cite{Galperin04,Frederiksen04,Solomon06,Rydnik} based on a parameterized
hamiltonian following Ref.~[\onlinecite{Monturet08}]. In spite of the hamiltonian
simplification, SCBA is numerically difficult and the evaluation of the spatial
distribution of the SCBA signal is beyond our capabilities, instead we have fixed the
STM-substrate symmetry and evaluated the IETS as a function of the applied bias.
Results are shown depending on temperature and modulation voltage of the experimental
lock-in amplifier in Sec.~\ref{result-iets}.

\section{Electronic structure of O$_2$ on A\lowercase{g}(110)\label{result-dft}}
% PDOS discussion

%%Figure 1
%
%%Table1

\begin{figure}
\suppressfloats
\includegraphics*[width=0.85\textwidth]{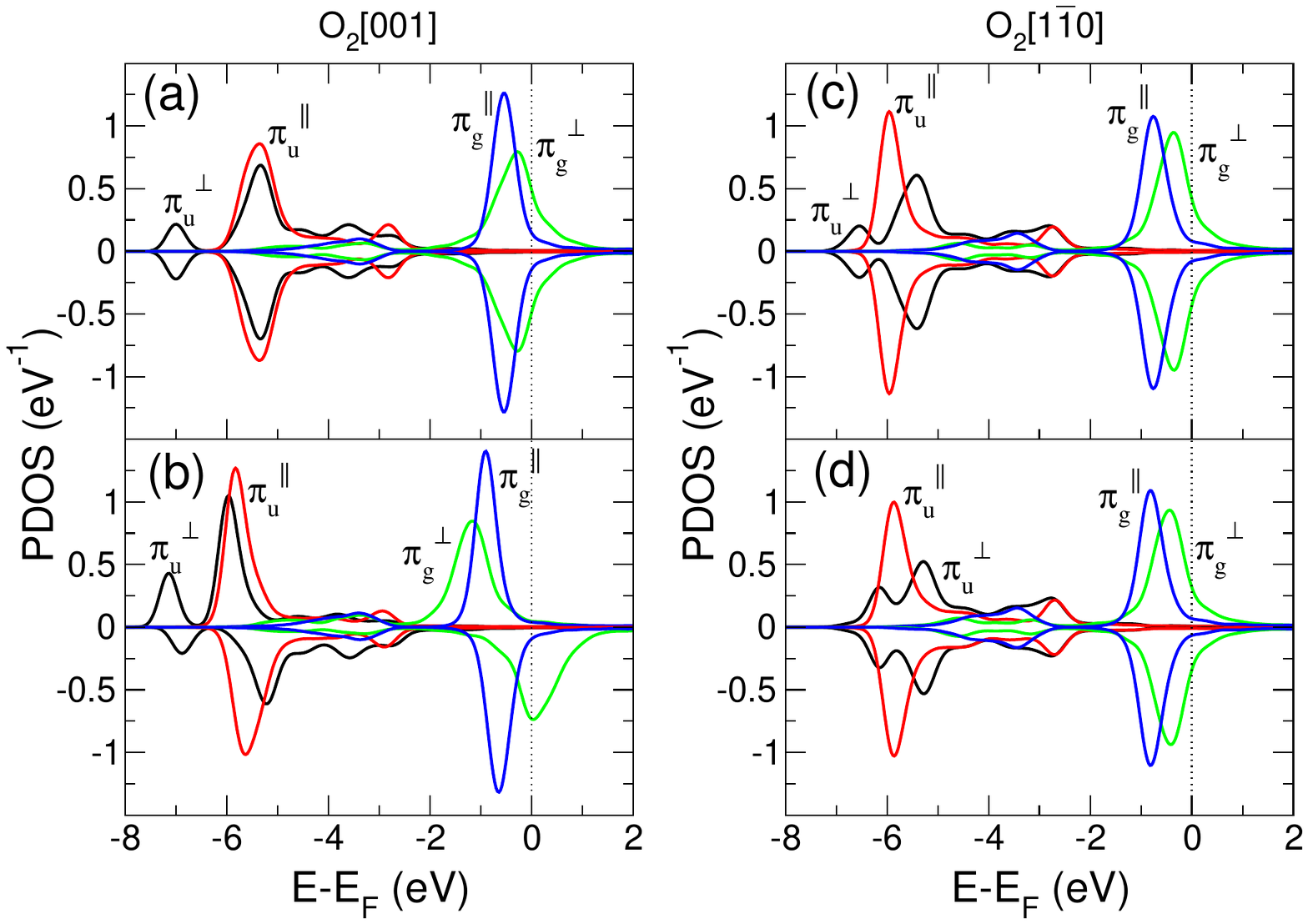}
\caption{ (Color online) Spin-up (positive) and spin-down (negative) projected density
of states of O$_2[001]$ (left panels) and O$_2[1\bar{1}0]$ (right panels) onto the
O$_2$ molecular orbitals. (a) and (c) Semilocal DFT (GGA) results. (b)and (d) Semilocal
DFT+$U$ results ($U=$3~eV). Intra-atomic correlation leads to a change in the $\pi_g$
resonance for the O$_2[001]$ only.  The Gaussian broadening used in the projections is
0.25~eV.\label{pdos}}
\end{figure}

Despite local and semilocal DFT have proven to be accurate in
determining variational ground state properties of the many-body
system such as the electron density, it cannot assure a reliable
treatment of the spectroscopic properties. In many cases, the
electronic structure is incorrect due to the inadequate description
of the on-site Coulomb interaction that favors fractional occupancies
of energetically close states. This is precisely what happens with
the electronic structure of the O$_2[001]$ chemisorption state
calculated with GGA. In Fig.~\ref{pdos}~(a) the projection of the
density of states (PDOS) on to the O$_2$ molecular orbitals (MOs)
reflects the partial occupation of the four $\pi_g$ orbitals, despite
the adsorbed molecule is capturing less than two electrons. These
unrealistic populations are possibly at the origin of the discrepancy
between the experimental IETS and the GGA-based simulations. To solve
this failure we have performed DFT+$U$ calculations of O$_2$
chemisorbed on Ag(110). The DFT+$U$ is aimed to correct the
electronic structure without perturbing the physical magnitudes for
which the GGA excels. A detail comparison of these quantities is
shown in Table~\ref{table1} for the two chemisorption states. The
good agreement obtained with the GGA values ensures the correct
description of the system.

The more realistic DFT+$U$ electronic structure shows that the
paramagnetic nature of O$_2$ is partially preserved when chemisorbed
along the [001] direction. As shown in Fig.~\ref{pdos}~(b), the
adequate description of the intramolecular Coulomb interaction breaks
the spin degeneracy imposed by GGA and allows the full occupation of
the two parallel-to-the-surface $\pi_g$ orbitals ($\pi_g^\parallel$)
and the spin-up perpendicular-to-the-surface $\pi_g$ ($\pi_g^\perp$).
Still the spin-down $\pi_g^\perp$ is partially occupied in order to
preserve the correct electron density ascribed to the molecule. As a
result, this molecular orbital is at resonance with the Fermi level.
The electronic structure of this chemisorption state can be viewed as
a Stoner-like process in which the molecule reduces its total energy
by splitting the orbitals according to their spin polarization. One
becomes completely filled at the expense of the other stabilizing the
full system.

In contrast, no spin-splitting of the $\pi_g^\perp$ orbital is
obtained for the O$_2[1\bar{1}0]$ molecule. At first sight there are
no significant differences on the structural properties
(Table~\ref{table1}) and the GGA electronic structures of the two
chemisorption states [compare Figs.~\ref{pdos}~(a) and (c)]. These
similarities are indeed consistent with the experimental observations
of almost identical chemisorption energies~\cite{bafrprl98} and equal
probability to find any of the two configurations~\cite{Hahn}.
However, the small differences found in the GGA calculations already
point to a slightly larger charge transfer from the surface to the
O$_2[1\bar{1}0]$ molecule, which is also closer to the surface and
more stretched~\cite{charge}.  Similar results were also found in
Ref.~[\onlinecite{Bird}] using a 3~$\times$~2 surface cell.
Experimentally, this picture agrees with the existence of two
energetically-close modes that are attributed to the O--O stretching
vibration of the O$_2[1\bar{1}0]$ (79.5~meV) and the O$_2[001]$
(85~meV)~\cite{bafrprl98}. Our DFT+$U$ calculations show that the
induced density for O$_2[1\bar{1}0]$ is larger and
the Coulomb repulsion less important.
%the larger induced density over the
%O$_2[1\bar{1}0]$ molecule that tries to capture almost two
%electrons
%\marginpar{\em \tiny Watch out:\\ 1.3-1.4$e$ in O$_2[001]$
%and 1.7-1.8$e$ in O$_2[1\bar{1}0]$...\\ too risky?} makes the
%on-site Coulomb repulsion less important,
%being its effect
%negligible in the electronic structure.
Interestingly, what the DFT+$U$ results highlight is a new difference
between the two chemisorption states, namely, the paramagnetic
character that is {\em only} preserved in the O$_2[001]$ molecule.

\begin{figure}
\suppressfloats
\includegraphics*[width=0.75\textwidth]{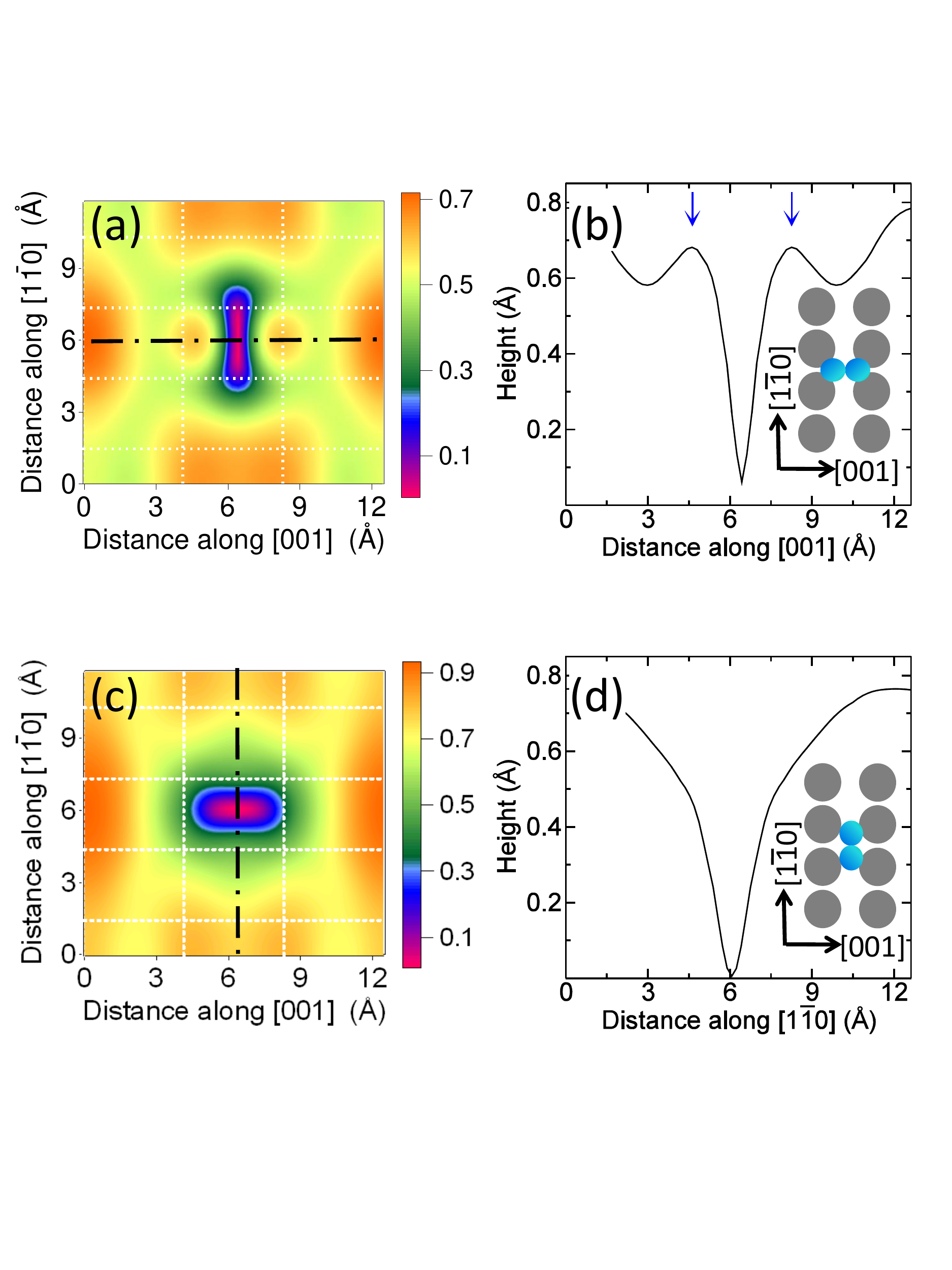}
\caption{ (Color online) Calculated DFT+$U$ ($U=$3~eV) local DOS for O$_2$ on Ag(110)
calculated with a sample bias $V$=200~mV. The zero height corresponds to a tip-surface
distance of $\sim$~6~\AA. (a) and (b) Topographical image and profile (both in \AA)
along the molecular axis for O$_2[001]$. The protrusions attributed to the
$\pi_g^\perp$ orbital are indicated by arrows. (c) and (d) Topographic image and
profile (both in \AA) along the molecular axis for O$_2[1\bar{1}0]$. The white grid
lines in (a) and (c) show the surface structure.\label{STM}}
\end{figure}

The DFT+$U$ electronic structures are used to generate the constant
current simulations of Fig.~\ref{STM}. The experimental topographical
images of the O$_2[001]$~\cite{Hahn} and the
O$_2[1\bar{1}0]$~\cite{hahoprl01} are well reproduced in both cases.
The main difference between both states are the two protrusions
appearing along the molecular axis in the former but not in the
latter. As was demonstrated by the non-spin polarized DFT
calculations of Olsson {\em et al.}~\cite{Olsson}, these protrusions
are the fingerprints of the $\pi_g^\perp$ orbital. The absence of
these features in the LDOS image of the O$_2[1\bar{1}0]$ molecule is
also understood on the basis of the stronger screening exerted on
this molecule by the surface electrons. Note, finally, that the
similar STM simulations obtained with non-spin polarized DFT and spin
polarized DFT+$U$ remark that information about the spin moment
cannot be directly extracted from the STM topography. In next
section, we will see that STM-based IETS can alternatively be used to
identify this peculiar electronic structure in which a spin-polarized
molecular orbital (MO) is at resonance with the Fermi level.

\section{Spatial distribution of constant current IETS
signals\label{result-sym}}

Our STM-IETS simulations, based on the DFT+$U$ electronic structure,
are able to reproduce qualitatively, and many times quantitatively,
the reported experimental findings: (i) the negative conductance
changes observed in the vibrational spectra of the O$_2[001]$
molecule, (ii) the spatial distribution of the conductance changes
recorded over the O$_2[001]$ molecule for the O--O stretch and the
antisymmetric O$_2$--Ag stretch vibrations, (iii) the absence of
signal for the center-of-mass and hindered rotation modes, and (iv)
the lack of IETS signal for the molecule chemisorbed along the
$[1\bar{1}0]$ direction~\cite{hahoprl01}.

\begin{figure}
\suppressfloats
\includegraphics*[width=0.75\textwidth]{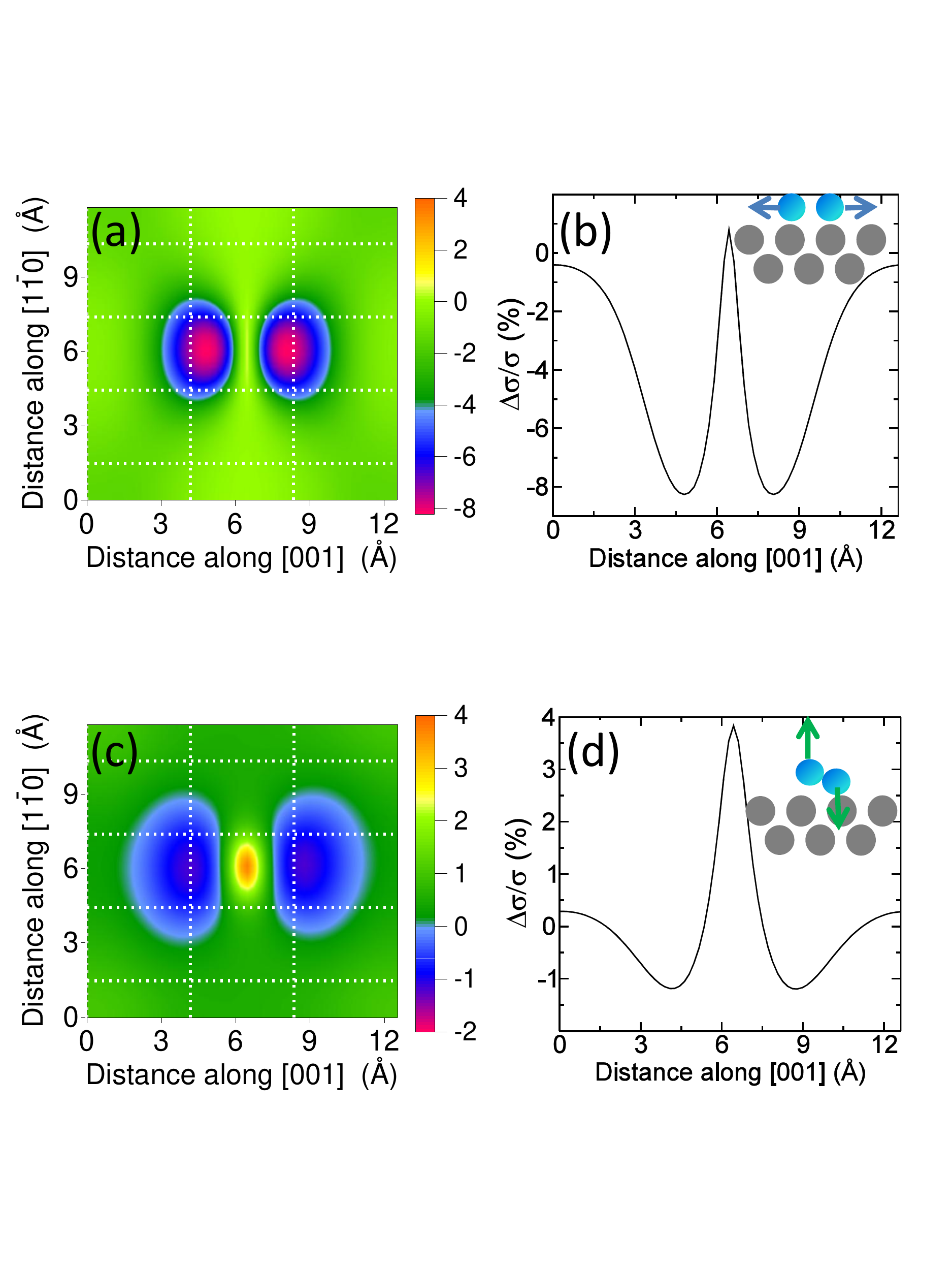}
\caption{ (Color online) Calculated DFT+$U$ ($U=$3~eV) STM-IETS for O$_2$ chemisorbed
on Ag(110) along the [001] direction. Spatial distribution of the conductance changes
for: (a) the symmetric O--O stretch mode ($V$=200~mV) and (b)the antisymmetric
O$_2$--Ag stretch mode ($V$=150~mV). (d) and (f) Cross sections of (c) and (e),
respectively, along the molecular axis. Note the negative and positive conductance
changes observed on the antisymmetric mode, in contrast to the overall negative values
of the O--O stretch mode. The white grid lines show the surface structure.
\label{IETS}}
\end{figure}

\begin{figure}
\suppressfloats
%\includegraphics*[width=0.45\textwidth]{IETS.eps}
%{\rotatebox{270}{\includegraphics*[width=0.55\textwidth]{figure4.prb.eps}}}
\includegraphics*[width=1.0\textwidth]{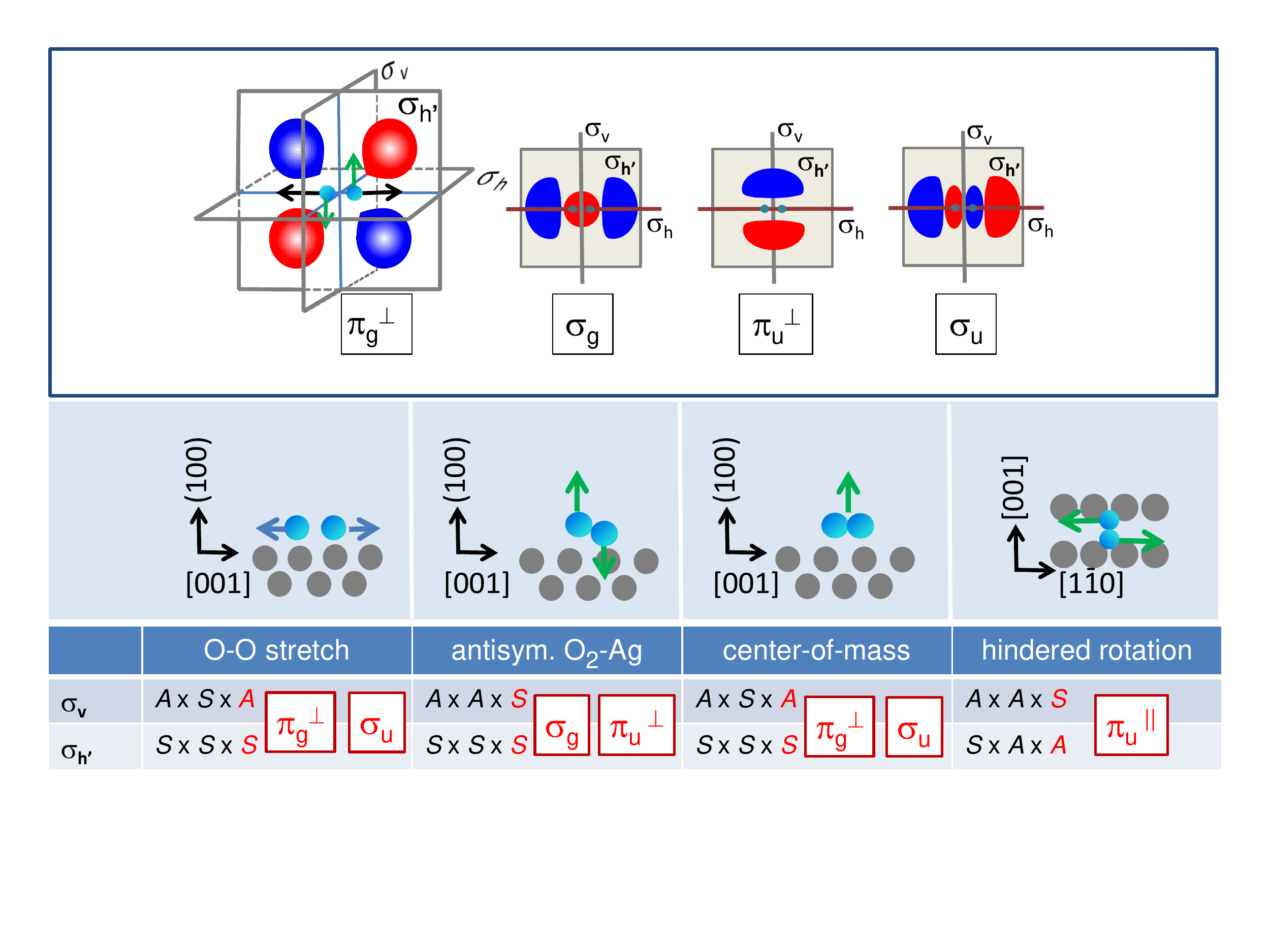}
\caption{(Color online) Sketch and symmetries of the MOs and modes involved in the IETS
images. The table summarizes the MOs that can couple with the $\pi_g^\perp$ to assure a
nonzero matrix element $\left \langle \psi_{m, \mathbf{k}}|v|\psi_{n, \mathbf{k}}\right
\rangle$ [see Eqs.~(\ref{ine}) and~(\ref{ela})]. The symmetric ($S$) or antisymmetric
($A$) character of the $\pi_g^\perp$ and the electron-vibration coupling $v$ is written
in black following this order for each vibrational mode and symmetry plane. In gray
(red), the symmetry character of $\psi_{m, \mathbf{k}}$ necessary for a nonzero
coupling. The MOs fulfilling those symmetry conditions are written inside the gray
(red) squares.
%Molecular symmetry elements
%and the corresponding symmetry of modes and molecular orbitals.
\label{SYM}}
\end{figure}

We start by analyzing the spatial distribution of the conductance
change for the O$_2[001]$ molecule. A schematic diagram of the
vibrational modes is depicted in Fig.~\ref{SYM}. Figures~\ref{IETS}
(a) and (b) show that the symmetrical O--O stretch mode leads to a
decrease of conductance over the adsorbed molecule. In agreement with
the experiment, the maximum absolute value of conductance change is
displaced from the center of the oxygen atoms in the same way as the
constant current STM image. This is reminiscent of the $\pi_g^\perp$
orbital, showing that the conductance decrease due to this
vibrational mode can be understood within the framework of a single
level resonant with the Fermi energy. Model
calculations~\cite{Persson} demonstrated that the presence of a
substantial density of states at the Fermi energy of large molecular
character leads to an important IETS signal of negative sign, {\em i.
e.} a reduction of conductance over the vibrational threshold, as
observed here.
%More precisely, the
%condition to have the decrease in conductance is that the
%resonance width be larger than the molecular level to Fermi
%energy distance.

%At the same time, the requirement of vibrational excitation
%implies an important lifetime of the resonance~\cite{Monturet08}. Adding the large
%intra-atomic correlation of oxygen, we retrieve the conditions for the Stoner
%splitting, namely, a molecular level at a distance from the Fermi energy shorter than
%the molecular width and an intra-atomic correlation energy larger than the width.
%Hence, tunneling through a spin-polarized molecular level necessarily leads to a
%decrease of conductance. This is precisely what we find.

The antisymmetric stretch mode is more interesting. Here, the oxygen
atoms move vertical to the surface with opposite phases. Again, in
good agreement with the experiment, we find that while the change in
conductance is negative away from the molecular center, it is
positive at the molecular center [Figs.~\ref{IETS} (c) and (d)]. In
contrast to the O--O stretch mode, this behavior implies that the
IETS signal involve the coupling with various molecular orbitals and,
therefore, we should go beyond the single-level model to understand
this IETS image.

% No signal for FR, CM
Simulations performed for the center-of-mass mode show a negative
conductance change image rather similar to the one obtained for the
O-O stretch mode.  The maximum conductance decrease however is a
factor of 5 lower in the center-of-mass, what explains the absence of
signal in the experiments for this mode. The simulated efficiency for
the hindered rotation mode, in which the atoms try to move opposite
to each other along the $[001]$ surface direction, is even smaller:
less than 0.8\% of conductance change. Neither is this mode observed
in the recorded IETS data.

%% Explaning the symmetry analysis and the multi-orbital model

All the above results can be understood with the help of
Eqs.~(\ref{ine})~and~(\ref{ela}). As the $\pi_g^\perp$ is the only MO
at resonance with the Fermi level, the topography of the conductance
changes is determined by the molecular orbitals coupling with such
single resonance, i.e., by those MOs contributing with nonzero
matrix elements in Eqs.~(\ref{ine})~and~(\ref{ela}). The symmetry
analysis summarized in the table of Fig.~\ref{SYM} provides a
preliminary idea of the MOs that can couple with the $\pi_g^\perp$
for each vibrational mode. The weight of these couplings in the IETS
image will ultimately depend on the orbitals energies and on the
strength of the electron-vibration coupling $v$, as discussed below.

There are two well defined symmetry planes in this system: the plane
perpendicular to the molecular axis $\sigma_v$ and the plane normal
to the surface that contains the molecular axis $\sigma_h'$. The
regions where the wave functions take positive and negative values
are schematically depicted for each MO in the upper panel of
Fig.~\ref{SYM} with blue and red colors, respectively. As written in
the table, the $\pi_g^\perp$ is antisymmetric ($A$) respect to
$\sigma_v$ and symmetric ($S$) respect to $\sigma_h'$. Since the O--O
stretch mode is symmetric respect to both planes, the coupling
electronic states should respectively be $A$ and $S$. These conditions
are only satisfied by the $\pi_g^\perp$ and the $\sigma_u$ orbitals. The same
conclusion is obtained for the center-of-mass mode, which is also
symmetric respect to both planes. The $A$ and $S$ characters of the
antisymmetric O$_2$--Ag stretch mode respect to $\sigma_v$ and
$\sigma_h'$, force the coupling of $\pi_g^\perp$ with the $\sigma_g$
and the $\pi_u^\perp$ orbitals. Finally, the excitation of the
hindered rotation mode, which presents a fully antisymmetric
character, couples the $\pi_g^\perp$ with the $\pi_u^\parallel$
orbital only.

%The change in conductance due to the SS mode is uniquely caused
%by the $\pi_g^\perp$. Both the inelastic and the elastic
%contributions shown in Figs.~\ref{SYM} (b) and (c),
%respectively, reflect this fact. The reduction of the
%conductance over the molecule is thus understood in the above
%framework of a single level resonant with the Fermi energy. The
%AS mode is more interesting. Here, the conductance change is due
%to the coupling between the $\pi_u^\perp$ and the
%$\pi_g^\perp$~\cite{energy}. The former rules the topography of
%the inelastic term [Fig.~\ref{SYM} (d)] and the latter that of
%the elastic one [Fig.~\ref{ELA-INE-1} (e)].}
%Indeed, at least two orbitals are contributing
%at the Fermi energy, the vertical $\pi_g^\perp$,
% and $\pi_u^\perp$. In order to have nonzero
%matrix elements, these orbitals couple with other orbitals that determine the
%topography of the change in conductance~\cite{Lorente2001}. For the decrease of
%conductance, the vertical $\pi_g^\perp$ is favored, again in agreement with the ...

\begin{figure}
\suppressfloats
\includegraphics*[width=0.75\textwidth]{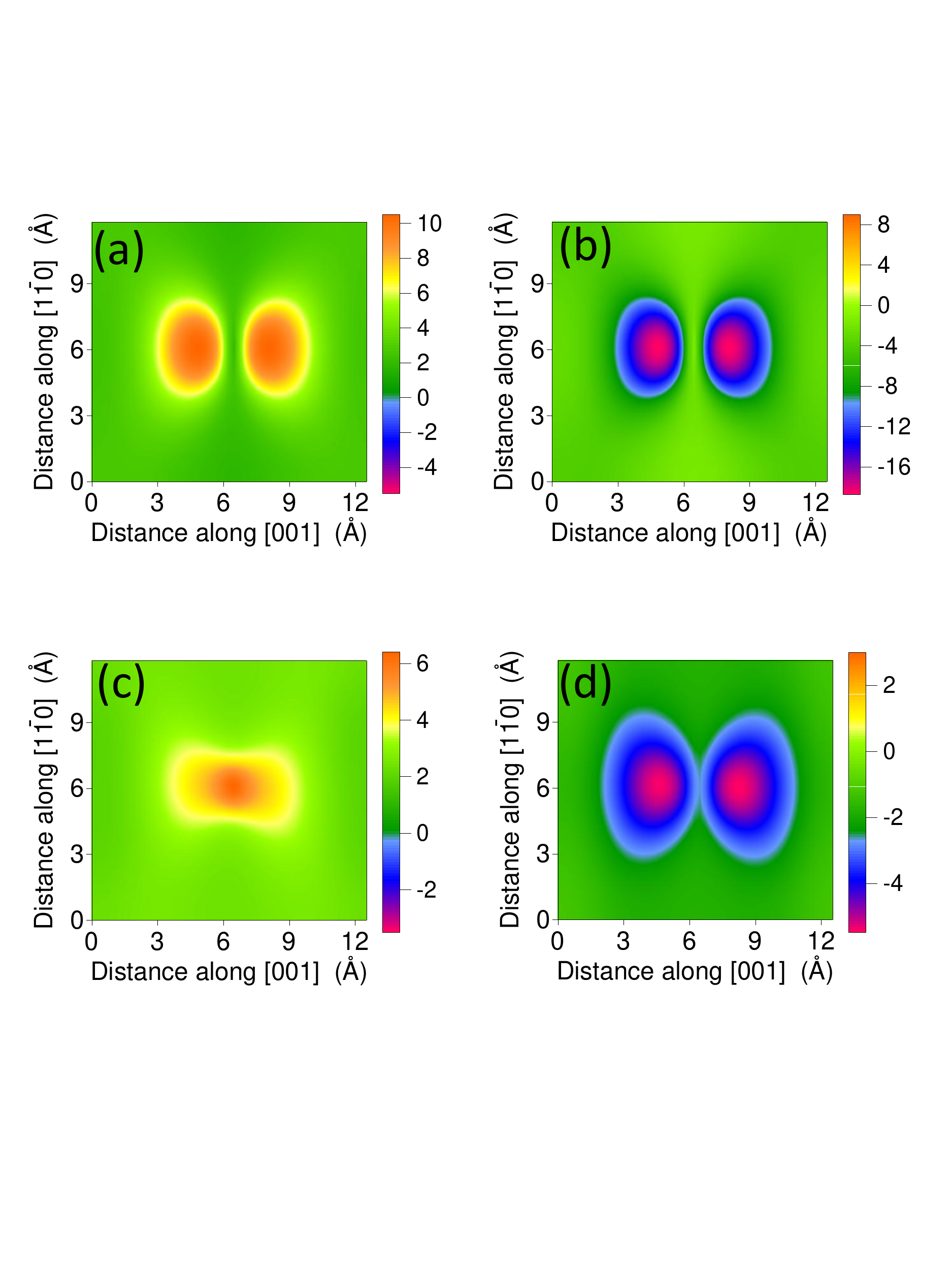}
\caption{(Color online) Contributions to the conductance changes shown in
Fig.~\ref{IETS}for O$_2[001]$. (a) Inelastic and (b) elastic contributions to the O--O
stretch mode. (c) Inelastic and (d) elastic to the antisymmetric O$_2$--Ag stretch
mode.
%Molecular symmetry elements
%and the corresponding symmetry of modes and molecular orbitals.
\label{ELA-INE-1}}
\end{figure}

\begin{figure}
\suppressfloats
\includegraphics*[width=0.75\textwidth]{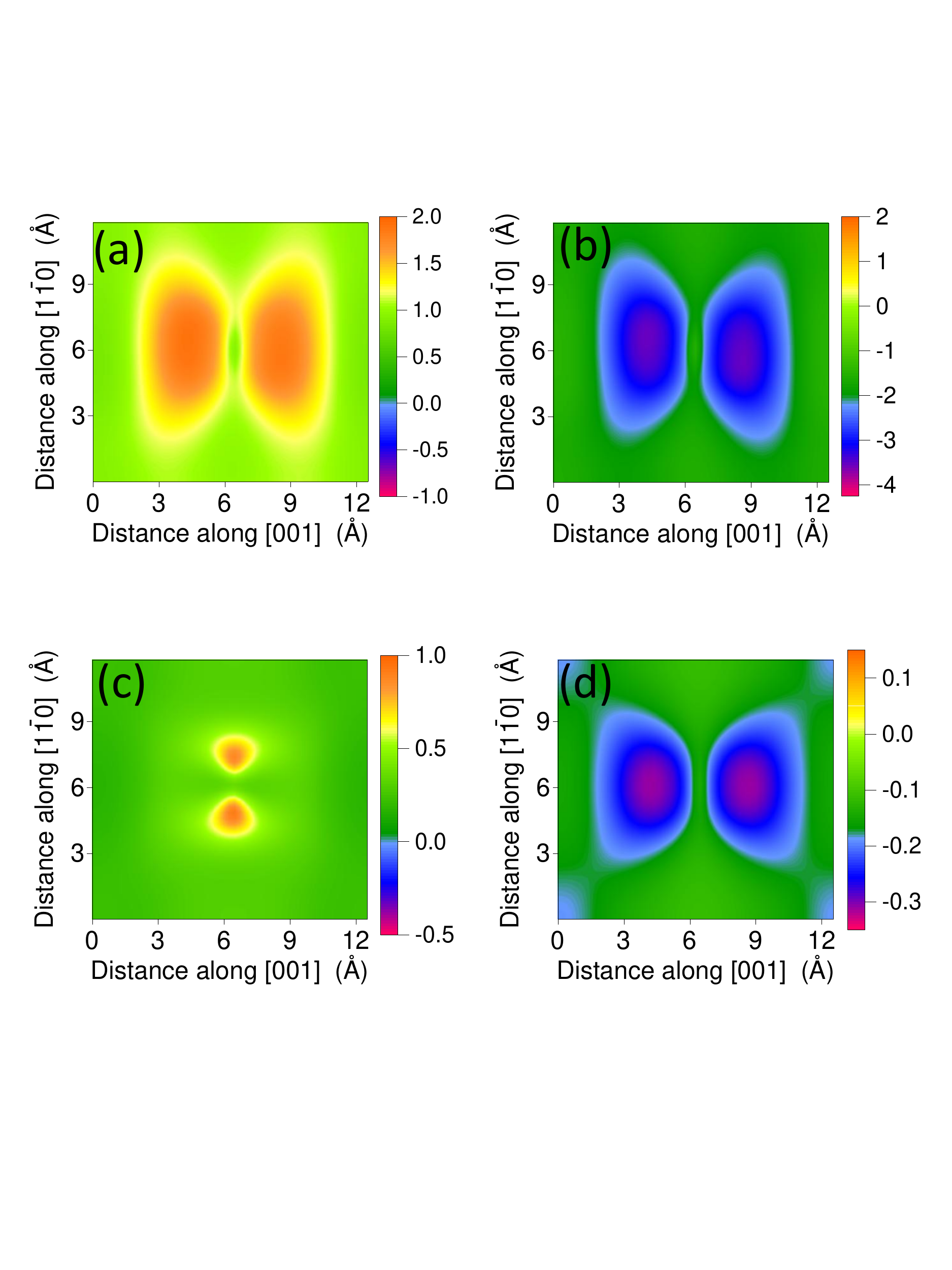}
\caption{(Color online) Contributions to the conductance changes for O$_2[001]$. (a)
Inelastic and (b) elastic contributions for the center-of-mass stretch mode. (c)
Inelastic and (d) elastic for the hindered rotation mode.
%Molecular symmetry elements
%and the corresponding symmetry of modes and molecular orbitals.
\label{ELA-INE-2}}
\end{figure}

Once we know which orbitals are coupled by each vibrational mode, the
topography of the IETS signal is better understood if the conductance
change is divided in its inelastic contribution due to the opening of
a new channel, hence positive, and the renormalization of the elastic
channel due to the presence of the vibration that leads to a negative
contribution. Figures~\ref{ELA-INE-1}~and~\ref{ELA-INE-2} show the
inelastic (left panels) and elastic (right panels) contributions for
the four vibrational modes. According to
Eqs.~(\ref{ine})~and~(\ref{ela}), the main difference between them
despite their sign, is that the inelastic part involves all possible
intermediate electronic states, while the states contributing to the
elastic part are restricted by the extra $\delta$-function to an
energy interval of the order of the mode frequency $\hbar \Omega$
about the Fermi energy. This strict energy restriction explains that
in all cases the $\pi_g^\perp$, being the only orbital at resonance
with the Fermi level, dominates the topography of the elastic
contribution.

In contrast, the inelastic image can be composed by all the coupling
states that according to the symmetry analysis contribute with
nonzero matrix elements in Eq.~(\ref{ine}). The weight of each
coupling state in the IETS image is partially determined by the
denominator of Eq.~(\ref{ine}), penalizing the orbitals that are
energetically distant from $E_F$. The effect of the different
weighting is reflected in the simulations of the inelastic
contributions. Figs.~\ref{ELA-INE-1}~(a) and \ref{ELA-INE-2}~(a) show
that the inelastic contribution is due to the $\pi_g^\perp$ (at
resonance with $E_F$) and not to the energetically distant $\sigma_u$
in the O--O stretch and center-of-mass modes. Otherwise, there would
be a certain intensity modulation along the molecular axis caused by
the nodes of the $\sigma_u$ orbital. Similarly, the absence of nodes
along the molecular axis and the maximum intensity obtained at the
center of the molecule in Fig.~\ref{ELA-INE-1} (c) show that the
image is basically due to the $\pi_u^\perp$ and not to the $\sigma_g$
in the antisymmetric stretch mode. In the hindered rotation mode,
Fig.~\ref{ELA-INE-2}~(c) shows that the inelastic contribution is
uniquely formed by the $\pi_u^\parallel$ as predicted by the previous
symmetry analysis.

%As a
%consequence, the inelastic contribution is due to the
%$\pi_g^\perp$ and not to the $\sigma_u$ in the O--O stretch and
%center-of-mass modes [see Figs.~\ref{ELA-INE-1}~(a) and
%\ref{ELA-INE-2}~(a)]. Similarly, the absence of nodes along the
%molecular axis and the maximum intensity obtained at the center of the molecule in
%Fig.~\ref{ELA-INE-1} (c) show that the image is basically due to
%the $\pi_u^\perp$ and not to the $\sigma_g$ in the antisymmetric
%stretch mode. In the hindered rotation mode,
%Fig.~\ref{ELA-INE-2}~(c) shows that the inelastic contribution
%is uniquely formed by the $\pi_u^\parallel$ as predicted by the symmetry
%analysis of Fig~\ref{SYM}.

Now we have all the ingredients necessary to understand the IETS
data recorded for the O$_2[001]$ molecule. In the O--O stretch
mode, the elastic and inelastic contributions are only due to
the $\pi_g^\perp$ orbital. As the elastic part is much larger
than the inelastic one, the excitation of this vibration leads
to an overall reduction in conduction except between the O atoms
where the $\pi_g$ orbital presents a nodal plane. Compared with
the O--O stretch mode, the smaller elastic and inelastic
contributions obtained in the center-of-mass mode are due to the
weaker electron-phonon coupling, as can be roughly inferred from
the different vibrational energy of each mode (see
table~\ref{table_modes}). In consequence, no
IETS signal is observed in the center-of-mass mode, despite its
symmetry similarities with the O--O stretch mode. The
antisymmetric O$_2$--Ag mode is more complex because involve
different orbitals: the inelastic part is dominated by
$\pi_u^\perp$ and the elastic part by the $\pi_g^\perp$. The
rich structure observed in this mode is a consequence of the
different nodal structure of the two orbitals. The total
conductance change is positive in the region between the two
oxygen atoms because the elastic contribution is zero whereas
the inelastic one is maximal. On the contrary, the conductance
change is negative out of the internuclear region because there
the elastic part is maximal and the inelastic part minimal. Finally, the
absence of IETS signal for the hindered rotation mode is a
consequence of two factors: the low DOS with $\pi_u^\parallel$
character around $E_F$ and the weak electron-phonon coupling of
this mode.
% (we calculate a vibrational energy of 33.8~meV for this
%mode\marginpar{\em \tiny Watch out:\\ Is it risky to write the
%theoretical vibrational energies?\\ Remember that they do not compare
%'well' with experiments...}).

We also perform simulations of the conductance changes for the
O$_2[1\bar{1}0]$ chemisorption state. The absence of
experimental IETS signal is in agreement with our calculations
that show changes of about 1\% at most. The reason behind this low
conductance changes is clear: as the $\pi_g^\perp$ is
practically occupied, the probability to excite any
vibrational mode by tunneling trough this MO is small.

%The low conductance changes
%are due to a reduced molecular density of states as the
%disappearance of the molecular spin and the total occupation
%of the $\pi_g^\perp$ orbital signal.

\section{Bias dependence of IETS: a self-consistent Born approximation study\label{result-iets}}

In order to compute the bias dependence of the inelastic signal, we
use the self-consistent Born approximation SCBA. Briefly, the SCBA
consists in using the Born expression for the electron-vibration
self-energy (lowest order in the electron-vibration matrix element)
replacing the free-electron Green's function by the full Green's
function~\cite{Jauho}. This last function is unknown, hence an
iterative process is used where the starting step is the
free-electron Green's function, and the second step uses the
self-energy computed with the Green's function obtained from the
first step. This self-consistent procedure is stable and has the
virtue of generating high-order terms in the electron-vibration
coupling.

The SCBA has been implemented in DFT based
approaches~\cite{Reimers}. But it is a costly method for
evaluating STM scans as the ones shown herein. It is however feasible
for a single STM conformation, where the bias is ramped and a
$d^2I/dV^2$ spectrum is obtained. Here, we have simplified the
problem and taken the tight-binding parametrization of
Ref.~\onlinecite{Monturet08}.

As discussed in Ref.~\onlinecite{Ness}, the SCBA fails in the limit
of large electron-vibration coupling. The main reason is because the
SCBA fails in taking into account the correct contribution of the
number of excited modes. Indeed, H. Ness~\cite{Ness} shows that SCBA
yields the same result as an exact calculation if the
electron-vibration matrix elements did not depend on the number of
excited modes. If the average number of excited modes is in the order
of one, then the SCBA gives a very good account of the vibrational
excitation even in the case of extraordinary vibronic
effects~\cite{Frederiksen08}. Approaches going beyond SCBA are
available~\cite{Galperin06,Thoss} and are necessary for exploring
realistic situation where the electron-vibration coupling is strong
enough to excite several overtones.

\begin{figure}
\suppressfloats
%\begin{center}
%\includegraphics*[width=0.75\columnwidth]{figure7.prb.eps}
\includegraphics*[width=0.75\columnwidth]{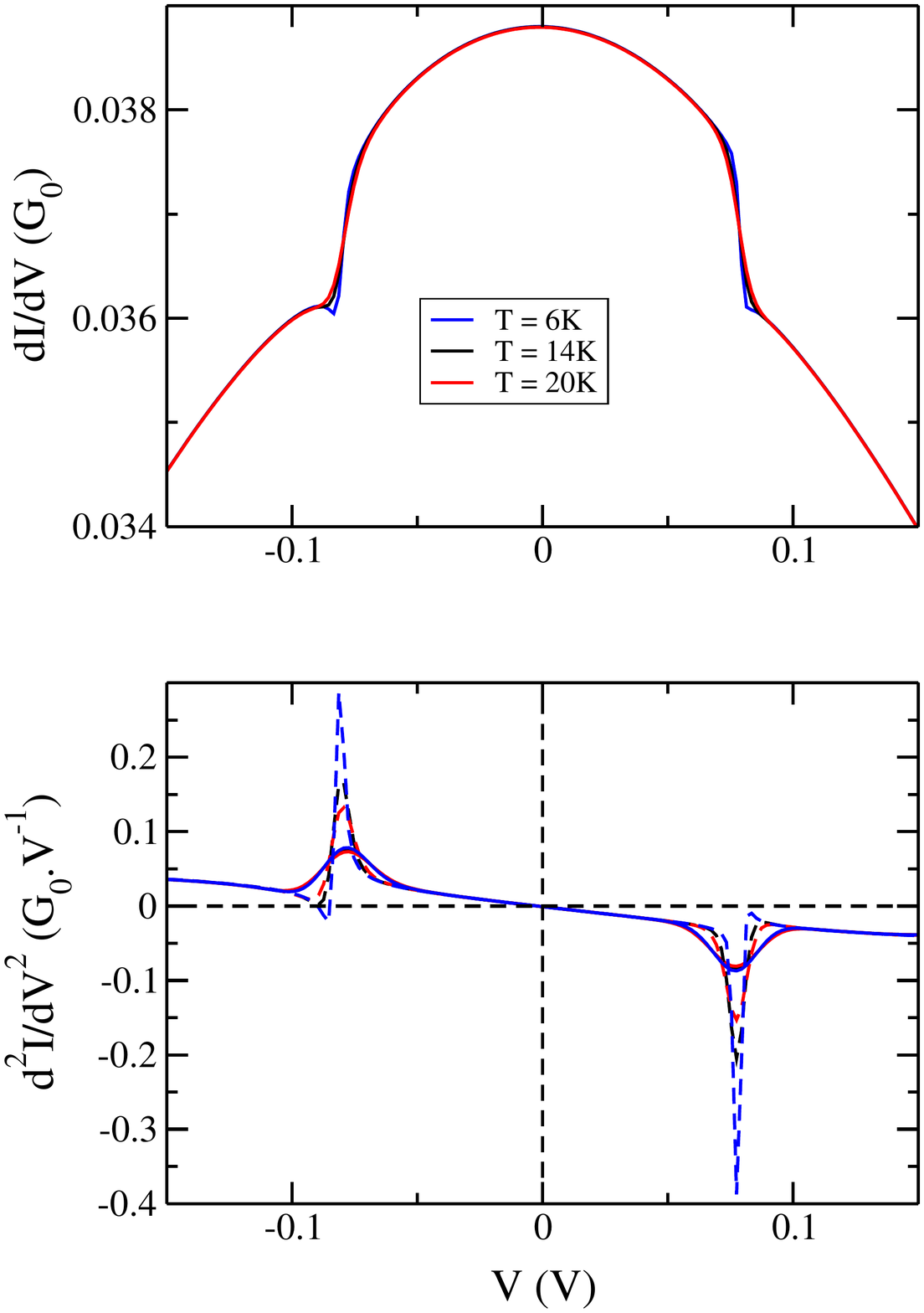}
%\end{center}
\caption{(a) Conductance of O$_2$ on Ag (110) along the [001] direction in a model
tunnel junction. The O--O stretch mode is allowed to interact with the tunneling
electrons leading to a decrease of conductance over the vibrational threshold. (b)
Derivative of the conductance with respect to the applied bias. Both graphs show the
behavior with temperature. In (b) the full line spectra have been convoluted with a
gaussian broadening representing the effect of a 7 mV rms modulation voltage. (a) and
dashed lines in (b) are without modulation-voltage broadening. } \label{Fig7}
\end{figure}

Figure~\ref{Fig7} shows the results of the SCBA for the O--O stretch
mode. Panel (a) shows the conductance, mainly given by the molecular
resonance about the Fermi energy, and two sharp Stokes and
anti-Stokes changes in conductance. In (b) the conductance derivative
or $d^2I/dV^2$ is shown. In both cases three different temperatures
are used, 6 K, 14 K and 20 K. At low temperatures the $d^2I/dV^2$
shows a Fano profile as has already been discussed for the excitation
of vibrations by electronic currents~\cite{Galperin04}, however as
the temperature increases, the peak becomes broadened and closer to a
simple Lorentzian peak. It is the modulation voltage used in the
lock-in detection that finally broadens and removes any particular
detail from the excitation peak. Figure~\ref{Fig7} (b) shows in full
line the peaks after convolution~\cite{Frederiksen07b} using a
modulation voltage of 7 mV rms as in the experiment~\cite{Hahn}.

% CONCLUSIONS

\section{Conclusions\label{con}}

We have conducted STM-IETS simulations, based on the DFT+$U$ electronic
structure and applying an extension of the Tersoff-Hamman
theory to the presence of inelastic
effects~\cite{Lorente2000,Lorente2004}. Our results are able
to reproduce the experimental IETS recorded by Hahn {\em et
al.}~\cite{Hahn,hahoprl01}: (i)
the negative conductance changes observed in the vibrational
spectra of the O$_2[001]$ molecule, (ii) the spatial
distribution of the conductance changes recorded over the
O$_2[001]$ molecule for the O--O stretch and the antisymmetric
O$_2$--Ag stretch vibrations, (iii) the absence of signal for
the center-of-mass and hindered rotation modes, and (iv) the
lack of IETS signal for the molecule chemisorbed along the
$[1\bar{1}0]$ direction. These results permit
us to give an enhanced description of the IETS signals in
terms of the molecular symmetries of the adsorbed system. Leading
to the conclusion that a multi-orbital description is mandatory
and that the simulation is able to capture the complex increases
and decreases of conductance depending on the STM tip location
over the molecule.

The good agreement of the simulations with the experiment validates
the results of the electronic structure as computed with DFT+$U$.
In particular, DFT+$U$ shows that out of the two possible molecular
conformations on the surface, only the one along the $[001]$
direction is paramagnetic.

The inclusion of the $U$ correction does not alter the values
of the other DFT parameters such as the molecular geometry
and vibrational structure. Regarding vibrations, in order
to attain reliable numerical values of the molecular frequencies,
we have used a Morse potential to fit the DFT potential energy
surface because anharmonicities rapidly set in and good frequency
values are difficult to obtain.

Finally, Fano-like profiles are revealed for the IETS as a function of the bias
voltage, however the two main sources of experimental broadening (temperature and more
importantly the lock-in modulation voltage) erases any information of the actual
spectral line shapes.

\begin{acknowledgments}
We acknowledge financial support from the Spanish MICINN (No. FIS2007-066711-CO2-00 and
No. FIS2009-12721-C04-01), and the Basque Government - UPV/EHU (grant No. IT-366-07).
Computational resources were provided by the Centre de Calcul de Midi-Pyr\'en\'ees, the
DIPC and the SGI/IZO-SGIker.
\end{acknowledgments}

%%%%%%%%%%%%
%REFERENCES
%%%%%%%%%%%%

%Table1
\begin{table*}
\suppressfloats \caption{Mode frequencies (meV) obtained
by DFT+$U$ ($U$=3~eV) and diagonalization of the dynamical
matrix computed with the same displacement $\Delta R$ for
all atomic oxygen coordinates with O$_2$ on Ag(110) along the
$[001]$ direction. The modes are schematically depicted in
Fig.~\ref{SYM}.
%labelled O--O
%for the O--O stretch mode,
%$\nu_{a}$ for the antysimmetric O$_2$-Ag stretch, $CM$ for the
%center-of-mass or symmetric O$_2$-Ag stretch and $FR$
%for the hindered rotation parallel to the surface.
} \label{table_modes}
\begin{ruledtabular}
\begin{tabular}{ccccc}
Mode &$\Delta R=0.025$~\AA~& $\Delta R=0.03$~\AA~& $\Delta R=0.04$~\AA~& Experiment\\
\hline
O--O stretch &91.5&89.7&90.4&82.0~\cite{Hahn} \\
antisym. O$_2$-Ag &42.0&31.8&33.6&38.3~\cite{Hahn} \\
center-of-mass &35.2&22.5&25.1&30~\cite{Sexton,Backx}\\
hindered rotation &34.5&37.8&34.5&- \\
\end{tabular}
\end{ruledtabular}
\end{table*}

%Table2
\begin{table*}
\suppressfloats \caption{ GGA and DFT+$U$ ($U$=3~eV) quantities describing the physical
properties of the two chemisorption states of O$_2$ on Ag(110). } \label{table1}
\begin{ruledtabular}
\begin{tabular}{cccccccc}
& & & & O--O & O--Ag & O$_2$--surface & magnetization \\
 state& & calculation & & (\AA) & (\AA) & (\AA) & ($\mu_B$) \\
\hline
O$_2[001]$& & GGA & & 1.43 & 2.35 & 1.34 & 0.0 \\
& &GGA + $U$ = 3 eV & & 1.41 & 2.37 & 1.34 & 0.54 \\
\\
O$_2[1\bar{1}0]$& & GGA & & 1.45 & 2.36 & 1.12 & 0.0 \\
& &GGA + $U$ = 3 eV & & 1.48 & 2.35 & 1.12 & 0.0 \\
\end{tabular}
\end{ruledtabular}
\end{table*}
\clearpage
%Figure 0

%Figure 1
%

%Figure 2

%Figure 3

%Figure 6

%Figure 4

%Figure 5

%Figure 7

\newpage
\clearpage


\begin{thebibliography}{99}
\bibitem{energy} The coupling of the $\pi_g^\perp$ with the $\sigma_g$
is less efficient because of the large energy difference between the two
states.
\end{thebibliography}
\end{document}